# Black Holes and the Scientific Process

**Manojendu Choudhury**

**UM-DAE Centre for Excellence in Basic Sciences, Mumbai – 400098.**

Arguably, black hole is perhaps the most popular scientific term among the lay person. Perhaps it is the phrasing of the term 'black hole' which appeals to the popular imagination, offering some exotic visual of a cosmic object to the mind's eye. Although the prediction of such an object was first made in the late eighteenth century[1] it was in the sixties of the last century that the study of this class of objects entered the realm of mainstream research, chiefly because of two developments: first, the birth and rapid progress of X-ray astronomy, and second, the discovery of pulsars, a class of neutron stars which form the highly dense end point in the life cycle of a normal star. The direct evidence of the existence of neutron stars elicited the possibility of the existence of black holes too which has led to the vast diverse body of research work on this subject over the last few decades.

The difference between neutron star (or any other heavenly body, for that matter!) and black hole is very fundamental. The neutron star has a definite surface whereas for the black hole the nature of the surface is very ephemeral, as matter can not exist at the surface (or the boundary) of the black hole. This boundary of a non rotating black hole is known as the event horizon, as photon is unable to escape this boundary of the black hole. Hence, no matter can be observed inside the event horizon. Therefore this quantity is used as the boundary of the object, as well as a measure of the size of the black hole. For a rotating black hole the structure of the boundary is more complicated with the addition of more layers of surface to account for the angular momentum of the system.

## 1. Black Hole Mass and its Implications

The fact that even light can not escape the pull of gravity conjures up an image of extremely dense object. Although it is not possible to ascertain a well defined value of the density of a black hole, one may obtain a minimum density by dividing the mass of the black hole by the volume of a sphere whose radius has the dimension of the Schwarzschild radius, i.e. the event horizon of a non-spinning black hole, given by

$$R_{Sch} = \frac{2GM}{c^2},$$

where *G* is the gravitational constant and *M* is the mass of the black hole, while *c* is the velocity of light in vacuum. If the sun (with solar mass $M_\odot$) was to be compressed to a maximum volume defined by the $R_{Sch\odot}$, then the lowest density will be

$$\rho_{Sch\odot} = \frac{3c^6}{32\pi G^3 M_\odot^2}.$$

The numerical value of the above expression is $1.86 \times 10^{19} kgm^{-3}$, which is typically more than the nuclear density of atoms (for eg, the nuclear density of gold is $2 \times 10^{17} kgm^{-3}$). Therefore the idea percolates that black holes are the most dense objects found in this universe, especially so in the popular literature.

Let us examine this so called lowest density of black holes a bit more closely and quantify the values obtained under different scale of mass. For a stellar mass black hole to exist, the current theoretical models state that the mass of the compact object needs to be at least $3M_\odot$. Therefore for the smallest size black hole, the density, which scales as $\frac{1}{M^2}$ (see above), goes down to $2.07 \times 10^{18} kgm^{-3}$ while for a black hole with mass of ten solar mass the density obtained is two orders of magnitude less than $\rho_{Sch\odot}$ and is now comparable to the nuclear density.

The most massive black holes are found in the centre of galaxies (namely the active galactic nuclei – AGN, although it is believed that most galactic centres harbour a supermassive black hole including that of our own Milky Way[2]). The mass of these supermassive black holes lie in the range of $10^6 - 10^9 M_\odot$. Hence the density of the most massive black hole, typically, reaches a value of $18.6 kgm^{-3}$. Comparing this value with the densities of solid gold ($1.932 \times 10^3 kgm^{-3}$), water ($1 \times 10^3 kgm^{-3}$) and air ($1.23 kgm^{-3}$) we notice that the so called density of the supermassive black hole is actually not very high. In fact it is about two orders of magnitude less than the mean solar density ($1.408 \times 10^3 kgm^{-3}$).

Also, more importantly, at the mass scale of a supermassive black hole, there is no other object that can exist as a stable single body other than a black hole. The most massive of stars, the O type, even though it is million times smaller than the biggest black holes, would have gone through the whole life cycle under the burden of its own gravity in a few million years,. Hence, the picture that emerges is that at a scale of

$10^6 - 10^9 M_\odot$, the most compact objects that can exist have a density less than a typical star. Hence, the forces that govern the matter-radiation interaction at highest scale, ensure that the existence of the most massive of objects is availed with a loss of density. Hence, for extremely high scale of mass, the density has to fall even for the most compact objects, as the density scales with $\frac{1}{M^2}$ where $M$ is the mass of the compact object.

This simple exercise of calculating the average density of the different black holes illustrates the power of detailed quantitative calculation in forming the picture of physical scenarios. This approach is one of the pillars of scientific thinking on which the edifice of the scientific understanding is resting. Here, the quantitative estimates of the densities of the various sizes of black holes allows one to infer that the steep gravitational potential of the compact object is reduced as one increases the mass of the black hole, and by the time one reaches the most massive black holes observed the gravitational potential is far less steep than our own sun. But, despite the drastic reduction in the density and hence the resultant gravitational attraction of the compact object, light is unable to come out of the event horizon. This immediately suggests that the nature of spacetime continuum plays a much more important role in determining the ability of massive objects to escape the gravitational attraction of a compact object, which is actually a manifestation of the bending of the spacetime continuum in the presence of the mass of the compact object. This has important implications in the way the black holes impact their immediate neighbourhood, manifested in the emission from the accretion disc around it (see section 3).

The obvious caveat to the above argument is that the density of black hole is ill defined. Without a proper quantum theory of gravity it is not yet known what the nature of mass inside the event horizon is like. In the above calculations it was assumed that the mass is homogeneously distributed inside the event horizon filling up the Schwarzschild volume, which is an unqualified assumption to say the least. Specifically, neither the effect of the singularity on the constituents around it is known, nor can the effect of the changed nature of the spacetime continuum on matter inside the event horizon be predicted. Nevertheless, the mean value of the density obtained above does provide a physical picture at the various scales of the mass of the black holes that exist.

## 2. Light from Black Holes

It is obvious from the nomenclature that no light can come out of the black hole, rendering the object to be invisible, literally. This leads to the conundrum of observing a black hole. Hence, despite being a theoretical curiosity for nearly two centuries as well as a logical inference of Schwarzschild's solution to Einstein's field equations[3], the realm of black holes remained in the periphery of mainstream research. On the theoretical front the emission from black holes received new fillip when quantum analysis of the spacetime around event horizon showed that radiation with spectrum similar to that of black body is emitted by the event horizon[4]. This radiation is called the Hawking radiation and its spectrum is characterized by a temperature which is inversely proportional to the mass of the black hole. This briefly raised the hope that an external observer may receive high amount of radiation from black holes. But despite the development of the very exciting niche of black hole thermodynamics[5], the power emitted from the Hawking radiation remains abysmally small and below the ability of any detector in foreseeable future.

The classical understanding of the physical picture of black holes can be summarized by the 'no hair conjecture' which states, without a complete formal proof, that the matter when crossing the event horizon does so irreversibly such that no information about the matter can be leaked out[6][7]. This leads to the famous black hole information paradox which contradicts the quantum mechanics paradigm in which a wave function that defines a particle will describe it for all time, future and past, determined by a unitary operator. Even the ability of black holes to radiate via Hawking radiation, which is a quantum effect, was unable to resolve the paradox unequivocally. In fact, reportedly, there was a public bet with Kip Thorne and Stephen Hawking on one side and John Preskill on the opposite side, in which Thorne and Hawking claimed that general relativity would not allow the black hole to leak any information and the Hawking radiation originated outside the event horizon and hence would not violate the no-hair-conjecture. On the other hand, Preskill argued that the reversibility of the unitary operator that determines the wave function of the particles falling through the event horizon will alter the general relativistic understanding in a way that information will be not be lost. Currently the paradox is unresolved [8] although Hawking has rescinded his position and argues that the quantum fluctuation of the event horizon will cause the black hole to evaporate leading to leaking of information.

This scenario of theoretical understanding is a prime example of one of the most important methods of scientific thinking. In this mode of scientific development the theoretical understanding keeps on progressing by virtue of the derivation of mathematical results in the hope that some day in future the theoretical claims will hold up to the experimental and observational scrutiny. This line of theoretical development may lead to mathematical situation which can not be rendered a physical justification, and the inferential construct of such theoretical models need to be scrutinized and accepted by the community. Given the current scenario of disseminating information through informal channels, viz. blogs, online forums, social network and related media which can be manipulated to market opinions, the non-specialist needs to learn to apply discretion while accepting scientific claims. One needs to check whether proper rigorous peer scrutiny has been applied to the claims and whether the claims fall in the category of the scientifically acceptable, theoretically possible or that of a mere mathematical oddity.

A different mode of scientific process is more evident when it comes to the indirect observations that lay claim to a particular scientific construct like a black hole. The advent of X-ray astronomy, just before the era of Hawking and his radiation, opened the window to observe the matter falling into the black hole. Here it is the observations of unknown sources emitting radiation not hitherto observed that drive the exploration of new regimes of discovery while theoretical models are developed to construct a scientific rationale. Such a situation presented itself about half a century ago when the very first strong X-ray emission was observed from Galactic sources. The paradigm of black holes, neutron stars (and also white dwarfs) in a binary system with a normal main sequence companion was hypothesized to explain the origin of the high energy emission. In such a system the compact object, viz. the neutron star or the black hole, accretes matter from the binary companion which loses its gravitational energy which is emitted out in the form of electromagnetic radiation . In the case of AGNs, the supermassive black hole at the galactic centre accretes matter from its environment, predominantly consisting of inter-stellar material (ISM) and also the nearby stars and the matter being accreted loses its gravitational energy. This energy is converted by some physical process (see next section) into electromagnetic radiation. Thus, an indirect observation of black holes is made possible.

3. **X-ray Astronomy and Accretion Process**

An X-ray photon is a quantum of electromagnetic radiation with an energy, to an order of magnitude approximation, some 1000 times greater than that of optical photons. Traditionally, the soft X-ray band is defined as the energy range 0.5 – 12 keV (corresponding to wavelength of 25 – 1 Å), the hard X-ray extends to 50 keV and the energy range beyond it till a few MeV is regarded as soft gamma-rays, although this classification is not very stringent. X-ray astronomy pertains to the observation of the sky in this regime of the electromagnetic spectrum. The study of cosmic sources at these high energies of X-rays and soft gamma-rays began only in the early 1960's, after the serendipitous discovery of the low mass X-ray binary (LMXB) Sco X-1, which houses a neutron star (the compact object) and a low mass optical companion in the main sequence. The most common physical phenomenon that gives rise to emission in the X-ray band is accretion of matter onto compact objects. This process involves accumulation of diffuse gas or matter onto the compact object under the influence of gravity, and is expected to be responsible for the observed properties of a wide range of X-ray sources from X-ray binaries to AGNs.

A mass $m$ being accreted to a body of mass $M$ and radius $R$ will lose potential energy $\frac{GMm}{R}$, which, if converted to electromagnetic radiation, will cause the system to have a luminosity of
$$L = \frac{GMm}{R} = \frac{1}{2}mc^2\frac{R_{Sch}}{R} = \xi mc^2,$$
where $\xi = \frac{1}{2}\frac{R_{Sch}}{R}$ gives the efficiency of converting the gravitational energy dissipated and lost in the electromagnetic radiation via the accretion process. Thus the efficiency of energy conversion due to accretion simply depends upon the compactness of the accreting body. A white dwarf star with $M = M_\odot$, $R_{Sch} = 3 \times 10^4 m$ and $R \approx 5 \times 10^6 m$ corresponds to $\xi = 3 \times 10^{-4}$. Correspondingly for $1 M_\odot$ neutron star with $R = 15 km$ the efficiency $\xi = 0.1$ (the efficiency for nuclear burning in neutron stars is only $\xi = 0.007$), and for black holes $\xi$ ranges from 0.06 (Schwarzschild black hole) to 0.426 (maximally rotating Kerr black hole). Thus, black holes, particularly, maximally rotating ones, are the most powerful energy sources in the universe and accretion is the mechanism by which the energy is released. Despite this high efficiency of emission of electromagnetic energy, the balance between the outwardly directed radiation pressure (obtained from Thomson cross-section) and inwardly directed gravitational pressure limits the luminosity to a limiting value, called

*Eddington luminosity*, which is given by
$$L_E = \frac{2\pi R_{Sch} m_p c^3}{\sigma_T} = 1.3 \times 10^{31} \left(\frac{M}{M_\odot}\right) \text{ W.}$$
It should be noted that this is derived assuming spherically symmetric geometry. Thus, for a stellar mass black hole with, typically, a mass of $10 M_\odot$ the maximum luminosity observed should be of the order of $10^{39} erg/s$, whereas for the AGNs ($M \sim 10^6 - 10^9 M_\odot$) that limit is $10^{44} - 10^{47} erg/s$. It is possible to exceed the Eddington limit by adopting a different geometry, but not by a large factor. Further, the limit applies to steady-state situations and in none steady-state situations like supernova or the dwarf novae explosions the Eddington limit can be exceeded by a large margin.

The different densities of the black hole systems manifests in different steepness of the gravitational potential of the compact object. This results in the accreting matter giving rise to different spectra such that the thermal emission of the accreting matter peaks at different wavelengths. For the stellar mass X-ray binary systems the temperature peaks in the X-ray band while for the less steep gravitational potential of the supermassive black holes the spectra peaks in the ultraviolet (or at times the optical) band.

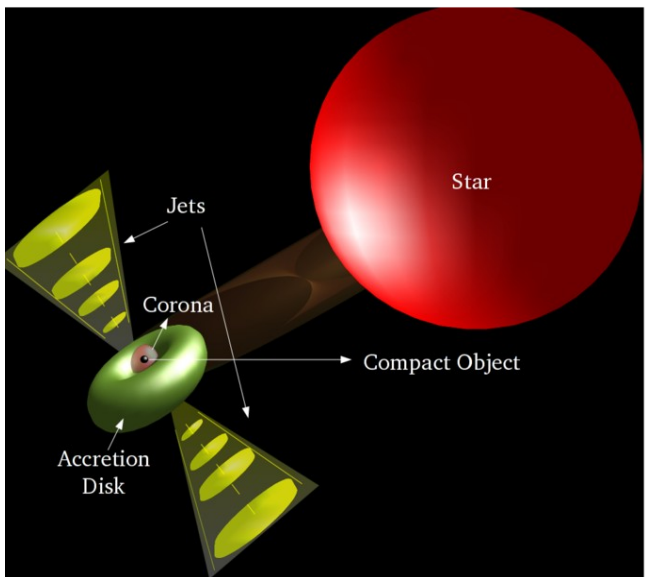

Figure 1. An artist's image of a X-ray binary system. (The image is reproduced under the conditions of the GNU Free Documentation License).

An additional interesting observational feature seen in these sources is that accreted matter is more often than not ejected out in the form of a jet perpendicular to the accretion disc. In fact the observations in the radio band, of outflowing conical jets from the core of the galaxies, often exhibiting superluminal (with apparent velocity greater than that of light) expansion, were reported before the phenomenon of accretion was inferred from the observational studies. Historically this happened for the extra-Galactic sources. Quasars, a sub-class of AGNs, were discovered in the radio band of electromagnetic radiation during the era of the very early discovery of X-ray sources which were subsequently identified to be accreting supermassive black holes ($10^6 - 10^9 M_\odot$) with outflows in the form of a jet observable in the radio band by virtue of the physical process of synchrotron emission. Therefore a paradigm of accretion being related to ejection had gradually developed, although the outflow was not observable in the radio quiet AGNs. The discovery of Galactic X-ray binaries exhibiting (superluminal) radio jets, with both physical and temporal (variability) scale roughly at 6 orders of magnitude less than those of quasars, led to the notion of ubiquitous presence of outflow in the form of collimated jets in accreting black hole systems and low magnetic field ($< 10^9 G$) neutron stars, lending them the terminology of microquasars.

The spectral analysis of the emissions from these sources show that in addition to the thermal emission from the infalling matter (normally following the geometrical form of a disc, hence the terminology of accretion disc) a highly energetic optically thin region also exists inside the truncated accretion disc giving rise to high energy non-thermal emission. Hence, the outskirts of the black holes provide the unique laboratory in the universe where matter-radiation interaction at the highest energies is observable.

The physical processes that cause the electromagnetic radiation from these systems involve the acceleration (positive or negative) of the charged particles due to the presence of electric and/or magnetic fields. The processes are, namely, Bremsstrahlung, Synchrotron, in addition to the scattering process of Comptonization. The emission behaviour, manifested in the emission spectrum, depends on the nature of the electron distribution in the accretion medium. Thus, the thermal and/or the non-thermal distribution of the electrons are of paramount importance in the emission process. It may be noted that the thermal black body emission is one of the most important component in the spectra of the emitted light from these accreting system and is used as a diagnostic of the effective temperature of the system. Also, it needs to be noted that the process of synchrotron, where the electrons are accelerated by the magnetic field, is observed in the emission from the outflow of the system.

## 4. Superluminal Motion

Among the various different physical phenomena seen in these sources, we will mention the most interesting one, namely, the superluminal motion. Due to the geometrical effects at high velocity, the outflowing jet appears to be moving at a velocity greater than that of light[9]. This optical illusion is rarely visible as it requires the combination of energetics of the system and the orientation of the expanding jet very near the line of sight. The extreme velocity is possible only if the central engine driving the system has the high compactness at the given scale of mass. The interesting aspect to be considered is that the features of the superluminal motion remains similar for both the extremes of black holes, stellar massive and supermassive.

An enthusiast, seeking to learn more of the observed studies of the phenomenon of superluminal motion, may follow two sources, 3C273 (AGN) and GRS 1915+105 (Galactic X-ray binary). Both these sources exhibit superluminal motion, and the time scale of the motion of the material on the plane of sight, is an indicator of the relative scales of the mass of the two objects. In Figure 2 the separation of the ejected blobs of GRS 1915+105 is measured over a few days, whereas for 3C273 the separation measured over much lower scale is measured over few years[10].

## 5. Scientific Process

In this article we have explored the power of the scientific process which not only enables us to explore invisible objects like the black hole but also extends our understanding in other spheres related to the core topic. Perhaps the most important aspect of the scientific process is to maintain the balance and consistency of the scientific construct, which the philosopher may refer to as harmony. Thus, the scientific process succeeded in converting an exotic theoretical object into a regular research material, and in the process created a connection of physical phenomena across a physical dimension spanning six orders of magnitude. This search for balance, consistency and harmony in the processes and constructs of the universe is not a very recent phenomenon, but what is fairly recent is the use of quantitative measurements in developing the understanding the of underlying process in a physical system. The first such successful attempt was that of Nicolas Copernicus's heliocentric model of the solar system[12]. As Thomas Kuhn puts it, "The sum of evidence drawn from harmony is nothing if not impressive. But it may well be nothing. "Harmony" seems a strange basis on which to argue for the Earth's motion, ... Copernicus's arguments are not pragmatic. They appeal, if at all, not to the utilitarian sense of the practicing astronomer but to his aesthetic sense and to that alone."[13] Copernican approach to explain the observation of the inferior planets and the retrograde motion of the superior planets produced a simpler model that seemed to predict the observations as opposed to Ptolemy's model which was only accommodating the observations by introducing inharmonious complexities in the system.

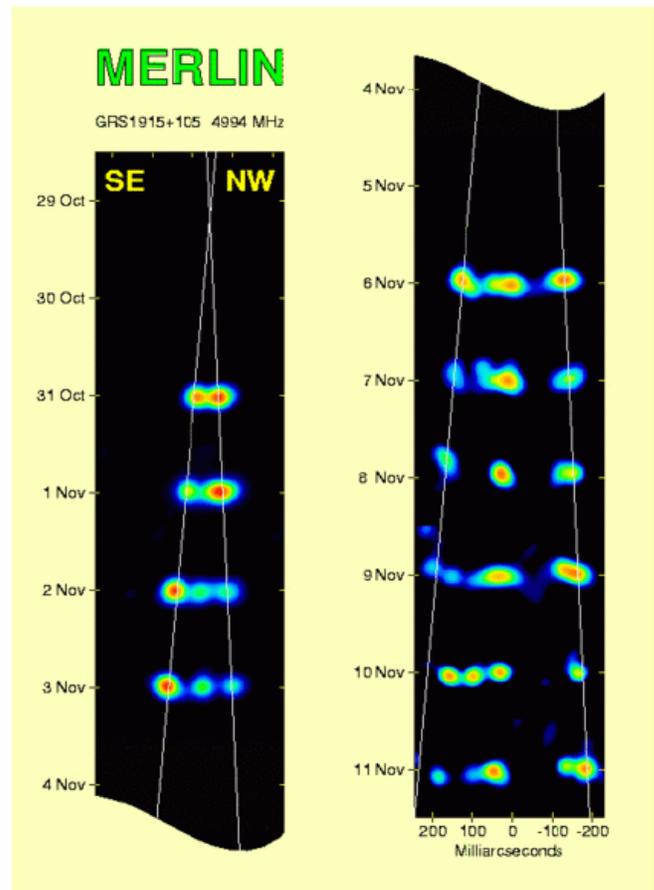

Figure 2. The observation of superluminal motion in the microquasar GRS 1915+105 from the MERLIN radio observatory. The two lobes appear to move apart with a velocity greater than that of light. (The image is produced under the Creative Commons 3.0 license[9]).

Similarly, the black hole is not without its controversies, in fact it is one of the most controversial topic of research since its introduction, but the scientific process of balancing our understanding consistently will continue to persevere and guide us to newer horizons of knowledge.